\documentclass[12pt,a4paper]{article}
\usepackage{graphicx,epsfig,epsf,rotating,fancyhdr,amsmath,natbib,subfigure}
\usepackage{txfonts}

\def\etal{{et al.}}

\def\3o{O~{\sc ii}}
\def\4o{O~{\sc iv}}

\title{\bf \vspace*{-15mm}A Mini-survey of Ultracool Dwarfs at 4.9 GHz}
\author{A. Antonova$^1$, J.G. Doyle$^1$, G. Hallinan$^2$, S.
Bourke$^2$, A. Golden$^2$} 
\date{$^1$Armagh Observatory, College Hill, Armagh BT61 9DG, N.
Ireland \\ $^2$ Computational Astrophysics Laboratory, I.T. Building,
National University of Ireland, Galway, Ireland}
\begin{document}
\maketitle

\abstract {A selection of ultracool dwarfs are known to be radio
active, with both gyrosynchrotron emission and the electron
cyclotron maser instability being given as likely emission
mechanisms. To explore whether ultracool dwarfs previously
undetected at 8.5 GHz may be detectable at a lower frequency. We
select a sample of fast rotating  ultracool dwarfs with no
detectable radio activity at 8.5 GHz, observing each of them at 4.9
GHz. From the 8 dwarfs in our sample, we detect emission from
2MASS J07464256+2000321, with a mean flux level of 286 $\pm$ 24 $\mu
Jy$. The light-curve of 2MASS J07464256+2000321, is dominated
towards the end of the observation by a very bright, $\approx $100
$\%$ left circularly polarized burst during which the flux reached
2.4 mJy. The burst was preceded by a raise in the level of activity,
with the average flux being $\approx$ 160 $\mu Jy$ in the first hour
of observation rising to $\approx$ 400 $\mu Jy$ in the 40 minutes
before the burst. During both periods, there is significant
variability. The detection of 100\% circular polarization in the
emission at 4.9 GHz points towards the electron cyclotron maser as
the emission mechanism. However, the observations at 4.9 GHz and 8.5 GHz
were not simultaneous, thus the actual fraction of dwarfs capable of
producing radio emission, as well as the fraction of those that show
periodic pulsations is still unclear, as indeed are the relative roles
played by the electron cyclotron maser instability versus gyrosynchrotron
emission, therefore we cannot assert if the previous non-detection at
8.5 GHz was due to a cut-off in emission between 4.9 and 8.4 GHz, or due
to long term variability.}
   
{\bf Keywords}Stars: low-mass, brown    dwarf -- Radio continuum: stars -- 
Radiation mechanism: general -- Masers

%

\section{Introduction}
Ultracool dwarfs (UCDs) are defined as those dwarfs with spectral
type M7 or later (Kirkpatrick \etal\ 1997). Due to the low and in most instances
non-detectable levels of H$\alpha$ and X-ray emission, radio
emission was considered to be insignificant until Berger \etal\
(2001) reported a detection from the brown dwarf LP 944-20. Additional
sources were reported by Berger (2002), Burgasser \& Putman (2005),
Berger (2006) and Phan-Bao \etal\ (2007) with the emission
mechanism assumed to be gyrosynchrotron. A highly polarized
flare, detected by Burgasser \& Putman (2005) from the M8 dwarf DENIS 1048-3956,
was interpreted as due to coherent electron cyclotron maser
emission.

Hallinan \etal\ (2006) reported a periodicity in the radio emission
of the UCD TVLM 513-46546 (hereafter TVLM 513), consistent with the
rotation period of the dwarf (Lane \etal\ 2007). They suggested that
the emission process was due to an electron cyclotron maser (ECM),
similar to the emission process in the magnetized planets in the
solar system (Zarka 1998, Ergun \etal\ 2000). In a higher
sensitivity followup study, Hallinan \etal\ (2007) observed
extremely bright, periodic bursts of both left and right hand
100$\%$ circularly polarized emission from the same source. The
characteristics of these short duration bursts were consistent with
a coherent process, the electron cyclotron maser instability.

There have been a number of surveys of ultracool dwarfs in the radio
(see references above). Thus far, 9 out of the $\approx100$ UCDs
observed at radio frequencies have been detected as radio sources.
All of these surveys, however, were conducted at 8.5 GHz. This may
have major implications for the number of radio active cool dwarfs
if the electron cyclotron maser is the dominant mechanism as in this
case the emission is mostly at the fundamental or second harmonic of the
cyclotron frequency $\nu_c \approx 2.8 \times 10^6 {\rm B [Hz]}$, i.e.
for an object to have detectable emission at 8.5 GHz requires a magnetic
field strength of $\approx$ 3 kG. Thus, it is possible that dwarfs, with
maximum field strengths below that value may be detectable in the
radio at lower frequencies.

In order to investigate this possibility, we conducted observations with
the Very Large Array (VLA)\footnote{The
National Radio Astronomy Observatory is a facility of the National
Science Foundation operated under cooperative agreement by
Associated Universities, Inc.}  for a sample of 8 ultracool dwarfs (previously
undetected at 8.5 GHz) at the lower frequency of 4.9 GHz. Here we
present the results of our
observations and the detection of another ultracool dwarf producing
electron cyclotron maser emission, the L0.5 binary system 2MASS
J07464256+2000321. 

\vspace*{-3mm}
\section{The Sample}
Our sample consists of 8 ultracool dwarfs spanning the spectral
range M8.5 -- T6. All our targets were previously undetected at 8.5
GHz. Below we summarize the main properties of each dwarf in the sample.\\

{\bf 2MASS J03350208+2342356 (2MASS0335+23)}\\
This object was identified as a young M8.5 brown dwarf, based on the
presence of Li~{\sc i} absorption, situated at a distance of $\approx $ 19.2 pc
(Gizis \etal\ 2000, Reid \etal\ 2002). It also exhibits H$\alpha$
emission, with $EW_{H\alpha}$ in the range 4.6 -- 6.5\AA, while
Berger (2006) reported an 8.5 GHz radio emission upper limit of $<
69~\mu$Jy. The projected rotational velocity $v \sin i$ = 30 km
s$^{-1}$ (Reid \etal\ 2002). \\

{\bf 2MASP J0345432+254023 (2MASS0345+25)}\\
This is an L0 spectral type brown dwarf (Kirkpatrick \etal\ 1999) with
$v \sin i \approx 25$ km~s$^{-1}$ (Berger, 2006 and references therein).
It shows no evidence of H$\alpha$ or Li~{\sc i} emission, has
$T_{eff} \approx$ 2430 K, log$ (L_{bol}/L_\odot) = -3.58 \pm 0.06$
(Vrba \etal\ 2004) at an estimated distance of 27 pc (Gizis \etal\
2003). Bailer-Jones \& Mundt (2001) reported it as photometrically
variable. The upper limit for the 8.5 GHz radio flux, given by
Berger (2002) is $< 88~\mu$Jy.\\

{\bf 2MASS J07464256+2000321 (2MASS0746+20)}\\
This dwarf was discovered by Kirkpatrick \etal\ (2000) and was later
resolved as a near equal mass binary system, with a separation of
2.7 AU (Reid \etal\ 2001). Dahn \etal\ (2002) determined its
distance to be 12.2 $\pm$ 0.05 pc. Vrba \etal\ (2004) estimated log$
(L_{bol}/L_\odot) = -3.64$ for the system, with an effective
temperature in the range $1900 < T_{eff} < 2225$ K. Several studies
reported detection of H$\alpha$ emission and only an upper limit on
Li~{\sc i} (Bouy \etal\ 2004). In the radio, Berger (2006) give an
upper limit of $F_{8.5GHz} < 48~\mu$Jy. Bouy \etal\ (2004) made the
first measurement of the dynamical mass of the system, classifying
both components as L0 $\pm$ 0.5 and L1.5 $\pm$ 0.5 respectively. In
a recent study, Gizis \& Reid (2006) argued that it is still unclear
whether 2MASS0746+20B is actually a brown dwarf. They argue that the
system is much older ($\gtrsim$ 1 Gyr) and the second component has
a mass which is above or just at the sub-stellar limit. Rotational
velocity studies give $v \sin i$ in the range 23.0 -- 28.8 km
s$^{-1}$ and a rotational period between 1.84 and 5.28 hrs (Bailer-Jones, 2004).
Such a rotational period is in very good
agreement with a photometric variability detected by Clarke
\etal\ (2002), with a periodicity of $\approx$ 3 hrs.\\

{\bf 2MASS J22244381-0158521 (2MASS2224-01)}\\
This is another fast rotator, e.g. Bailer-Jones (2004) reported a
projected rotational velocity in the range $20.9 <v \sin i < 29.2$
km s$^{-1}$ giving an expected rotational period of 3.8 hrs. The
dwarf was first discovered by Kirkpatrick \etal\ (2000), classifying
it as L4.5. The distance to the dwarf is 11.4 pc, it has $T_{eff}
\approx$ 1790 K (Dahn \etal\ 2002) and a bolometric luminosity log$
(L_{bol}/L_\odot) = -4.15 \pm 0.02$ (Golimowski \etal\ 2004).
H$\alpha$ emission has been detected, with $EW_{H\alpha}$ = 1\AA\
and only an upper limit for Li~{\sc i} (Kirkpatrick \etal\ 2000). In
the radio, Berger (2006) gives an upper limit of the flux at 8.5 GHz
$<
33~\mu$Jy.\\

{\bf 2MASS J15074769-1627386(2MASS1507-16)}\\
This is a nearby L5 brown dwarf, located at 7.3 pc (Dahn \etal\
2002). It has an estimated age of $\gtrsim$ 1 Gyr and $T_{eff}
\approx 1700$ K, log$ (L_{bol}/L_\odot) = - 4.27 \pm 0.05$ (Vrba
\etal\ 2004) and only upper limits on Li~{\sc i} and H$\alpha$ of
less than 0.1\AA\ and 0.5\AA\ respectively (Reid \etal\ 2000). For
this dwarf, Bailer-Jones (2004) reported a projected rotational
velocity of $v \sin i ~\approx 27$ km s$^{-1}$ giving an expected
rotational period of 3.5 hrs. Berger (2002)
give an upper limit to the 8.5 GHz radio flux as $< 36~\mu$Jy.\\

{\bf 2MASS J13054019-2541059 (Kelu-1)}\\
This is the fastest rotator in our sample with $v \sin i = 60 \pm 2$
km s$^{-1}$ (Mohanty \& Basri 2003), it has a parallax of 18.7 pc
(Dahn \etal\ 2002), and a bolometric luminosity of log$
(L_{bol}/L_\odot) = -3.59 \pm 0.04$, which together with an age,
constrained to 0.3 - 1 Gyr, yields $T_{eff} = 2100 - 2350$~K
(Golimowski \etal\ 2004 and references therein). Kelu-1 appears
over-luminous compared to other early type L dwarfs, which could be
either due to a very young age or a close companion (Leggett \etal
2002). It exhibits H$\alpha$ emission (Ruiz \etal\ 1997) but is
undetected in X-rays (Neuh{\"a}user \etal\ 1999) and radio
wavelengths. Berger (2006) gives an upper limit of $< 27~\mu$Jy
at 8.5 GHz. Using the Keck laser guide star adaptive optics system,
Liu \& Leggett (2005) resolved the dwarf as a binary with separation
5.4 AU, with estimated spectral types L1 - L3 and L3 - L4.5
respectively, with masses in the sub-stellar regime
for both components. Recently Audard \etal\ (2007) conducted
simultaneous X-ray and 8.5 GHz observations of this system, resulting
in a detection in X-rays, $L_X = 2.9 \times 10^{25}$ erg~s$^{-1}$
and a $3\sigma$ upper limit of $42 ~\mu$Jy at 8.5 GHz.\\

{\bf SDSS J162414.37+002915.6 (SDSS1624+00)}\\
This is the first field T dwarf discovered (Strauss \etal\ 1999),
and is classified as a T6. It has a parallax of 11.00 $\pm$ 0.15 pc
(Burgasser \etal\ 2006) and a bolometric luminosity log$
(L_{bol}/L_\odot) = -5.16 \pm 0.05$. It has a rotational velocity of
$v \sin i$ = 34 -- 38 km s$^{-1}$ (Zapatero Osorio \etal\ 2006). It
has no detected H$\alpha$ emission (Burgasser \etal\ 2000) and no
detection of radio emission at 8.5
GHz ($F_{8.5GHz} < 36~\mu$Jy,  Berger 2006).\\

{\bf 2MASP J1632291+190441 (2MASS1632+19)}\\
This is an L8 brown dwarf, first discovered by Kirkpatrick \etal\
(1999), situated at a distance of 15.2 pc (Dahn \etal\ 2002). It's
effective temperature and luminosity are $T_{eff}$ = 1346~K and
log$ (L/L_\odot) = - 4.6$ respectively (Vrba \etal\ 2004). This dwarf
has upper limits for emission in both H$\alpha$ (Mohanty \&
Basri 2003) and radio ($F_{8.5GHz} < 36~\mu$Jy, Berger 2006). The
reported projected rotational velocity for 2MASS1632+19 is $v \sin i = 30
\pm 10$ km~s$^{-1}$ according to Mohanty \& Basri (2003), while Zapatero
Osorio \etal\ (2006) give $20.9 \pm 7.0 < v \sin i < 22.6 \pm
6.7$ km~s$^{-1}$.
\begin{table*}
\caption{The sample of 8 ultracool dwarfs.}

\begin{tabular}{llllrrc}
Name & $\alpha$ (J2000) & $\delta$ (J2000) & SpT & $\pi$ & $v \sin i$ &
$F_{(4.9~GHz)}$ \\
 &  & &  & ($mas$) & (km s$^{-1}$) & ($\mu$Jy)\\ \hline

 2MASS 0335+23 & 03 35 02.08 & +23 42 35.6 & M8.5 & 51.2  & 30   & $<$ 45  \\
 2MASS 0345+25 & 03 45 43.16 & +25 40 23.3 & L0   & 37.1  &  25  & $<$ 36  \\
 2MASS 0746+20 & 07 46 42.56 & +20 00 32.2 & L0.5 & 81.9  & 25.8 & 286 $\pm$ 24  \\
 Kelu-1        & 13 05 40.18 & -25 41 06.0 & L2   & 53.6  & 60.0 & $<$ 84 \\
 2MASS 1507-16 & 15 07 47.69 & -16 27 38.6 & L5   & 136.4 & 27.2 & $<$ 57  \\
 SDSS 1624+00  & 16 24 14.37 & +00 29 15.6 & T6   & 91.5  & 36   & $<$ 81 \\
 2MASS 1632+19 & 16 32 29.11 & +19 04 40.7 & L8   & 65.6  & 30.0 & $<$ 39 \\
 2MASS 2224-01 & 22 24 43.81 & -01 58 52.1 & L4.5 & 88.1  & 24.7 & $<$ 46\\\hline
\end{tabular}

Note. - The columns are (left to right): name of the object; right
ascension; declination; spectral type; parallax; rotational
velocity; radio flux at 4.9 GHz  \  References. Ruiz \etal\ (1997),
Kirkpatrick \etal\ (1999), Strauss \etal\ (1999), Gizis \etal\
(2000), Kirkpatrick \etal\ (2000), Berger (2002), Dahn \etal\
(2002), Reid \etal\ (2002), Gizis \etal\ (2003), Mohanty \& Basri
(2003), Bailer-Jones (2004), Golimowski \etal\ (2004), Vrba \etal\
(2004), Berger (2006), Zapatero Osorio (2006).
\end{table*}

\vspace*{-3mm}
\section{Observations and Data Reduction}
The observations were conducted at a frequency of 4.9
GHz (6 cm) with the NRAO Very Large Array (VLA) on 22--23 January 2007. During the
observations, the instrument was in full array mode and DnC
configuration. We used the standard continuum mode with 2 $\times$ 50
contiguous bands, sampling every 10 s. The flux density calibrator
was 3C147. The phase was monitored using calibration sources selected
to be within 10 degrees of the target. The total time on each source
was 2 hours (time on source in a single scan being 8 minutes, before
moving to the phase calibrator for 90 s) enabling $3\sigma$ confirmation
of all sources with average flux levels of $ > 60 \mu$Jy.

Data reduction was carried out with the Astronomical Image
Processing System (AIPS) software package. The visibility data were
inspected for quality both before and after the standard calibration
procedures, and noisy points removed. For imaging the data, we
used the task IMAGR. We also CLEANed the region around each source
and used the UVSUB routine to subtract the resulting source models
for the background sources from the visibility data. We then used
the task UVFIX to shift the tangent point coordinates of the target
source to coincide with the phase centre of the map. The
subtraction of the background sources and the shifting of the map
are necessary since the side-lobes of those sources and the change
in the synthesised beam shape during the observation may result in
creating false variability of the target source, or contamination of any
real variability present. As a next step we re-imaged the visibility
data set in both total intensity (Stokes I), and circular
polarization (Stokes V). For examining the light curves we used the
AIPS task DFTPL.

\begin{figure}[hb!]
\vspace{9.5cm}
 \includegraphics{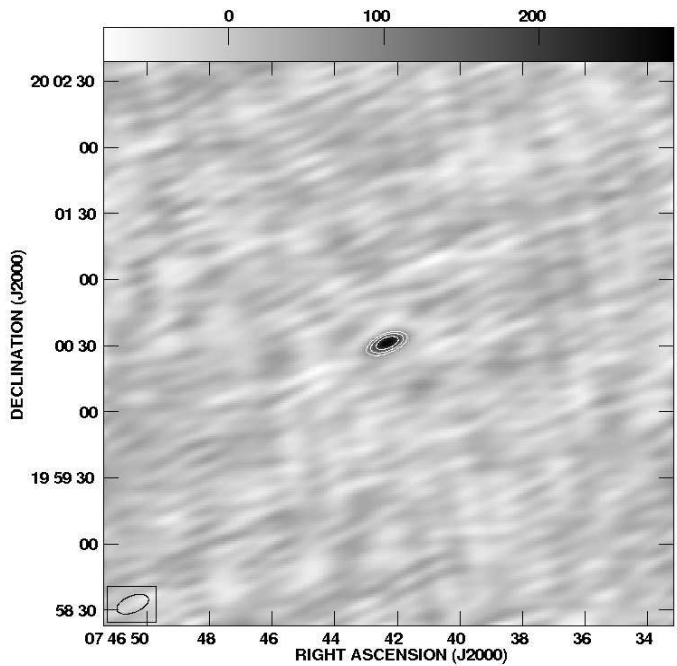}

\caption{A VLA map of 2MASS0746+20 at 4.9 GHz  with the background sources
removed. The image is in total intensity (Stokes I). Over-plotted are the
3, 5 and 8 $\sigma$ contour levels, where 1 $\sigma$ = 24 $\mu$Jy. The beam
size is shown at the bottom left corner of the map. \label{Fig1}}
\end{figure}

\section{Results}

In Table 1 we list the main properties of the observed sources, as
well as the measured radio flux or upper limits at 4.9 GHz, based on
detection/non-detection respectively. From the 8 dwarfs in our
sample, we detect emission from 2MASS0746+20, with a mean flux level
of 286 $\pm$ 24 $\mu Jy$ (Fig. 1). The object's derived position was
consistent with its expected location after correcting for proper motion
(Dahn \etal\ 2002).

The flux, detected from 2MASS0746+20 implies a radio luminosity
$L_{4.9GHz} =  5.09 ~\times 10^{13}$ erg $s^{-1} Hz^{-1}$ and
log($L_{R,\nu}/L_{bol}$) = --16.24. Due to
the low spatial resolution of our observation, we do not resolve the
system, thus we cannot determine which component is dominant.

To further examine the behaviour of the source, we plot its
light-curve in total intensity (Stokes I) and circular polarization
(Stokes V) (Fig. 2). We show  three different time resolutions
(10s,  30 s and 180 s)  The light-curve of 2MASS0746+20, is
dominated towards the end of the observation by a very bright,
$\approx $100 $\%$ left circularly polarized burst during which the
flux level reached 2.3 mJy. The burst was preceded by a raise in the
level of activity, with an average flux of $\approx$ 160 $\mu Jy$ in
the first hour of observation rising to $\approx$ 400 $\mu Jy$ in
the 40 minutes before the burst. This rise is in part due to a
series of other bursts, e.g. one at 4.3 hr. and another at 4.6 hr.
Unfortunately, the larger of these occurs during a phase
calibration, and thus we only detect the rise and delay. It is
however detectable in both I and V.

The above fluxes were determined by separately imaging the two time
intervals. During both intervals, there is significant variability.
Because the total observing time (2 hours) is less than the expected
rotational period of the dwarf ($\sim$ 3.7 hours, Bailer-Jones
2004), searching for rotational modulation (as in Hallinan \etal\
2006, 2007) is not possible.
\begin{figure*}[hbt!]
\vspace{12.0cm} \includegraphics{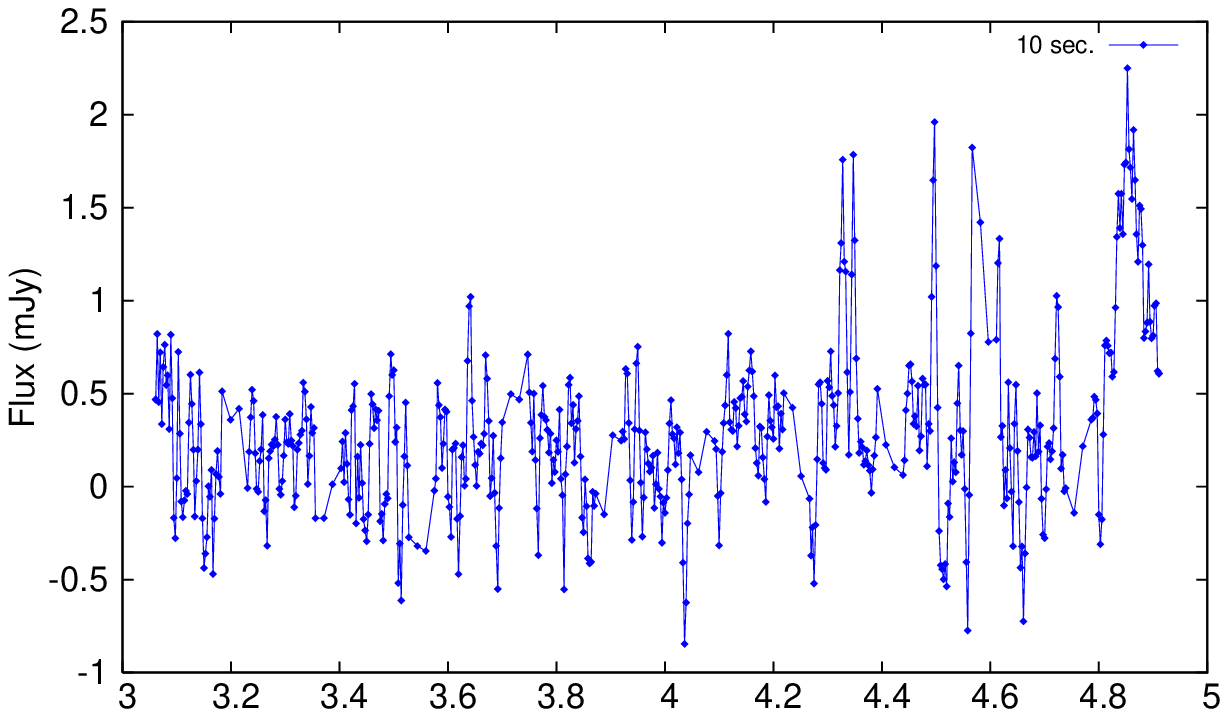}
\includegraphics{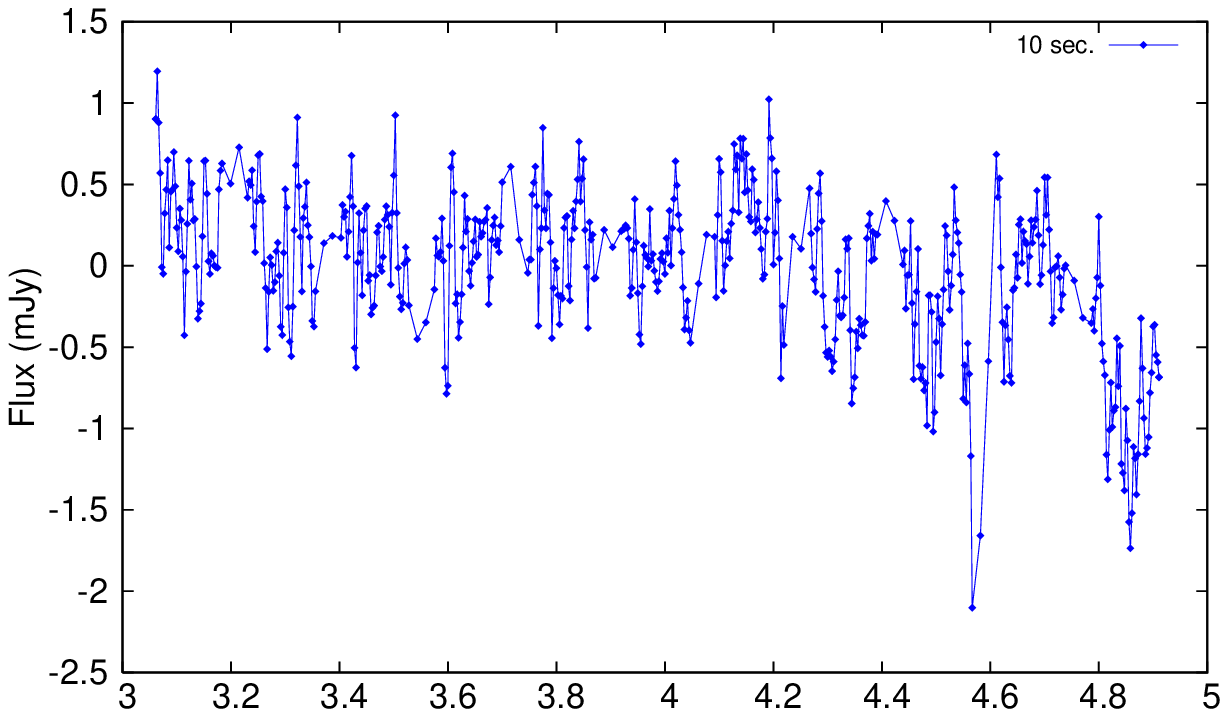} \vspace*{1.2cm}\includegraphics{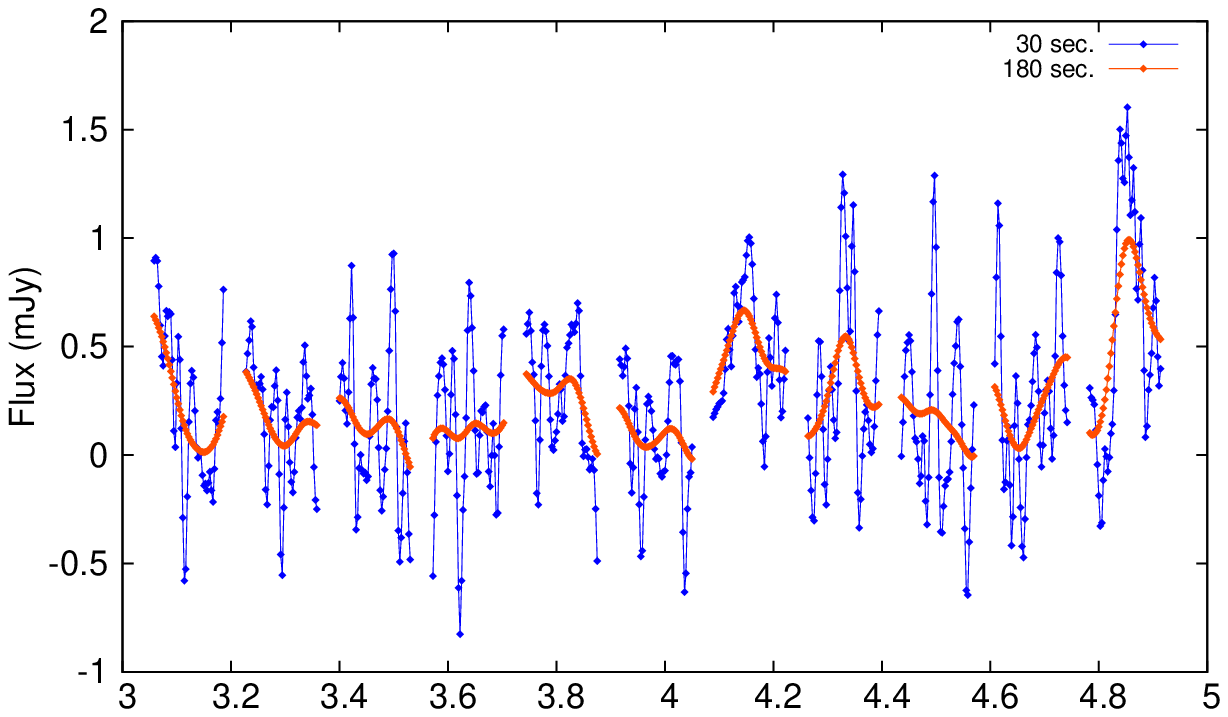}
\includegraphics{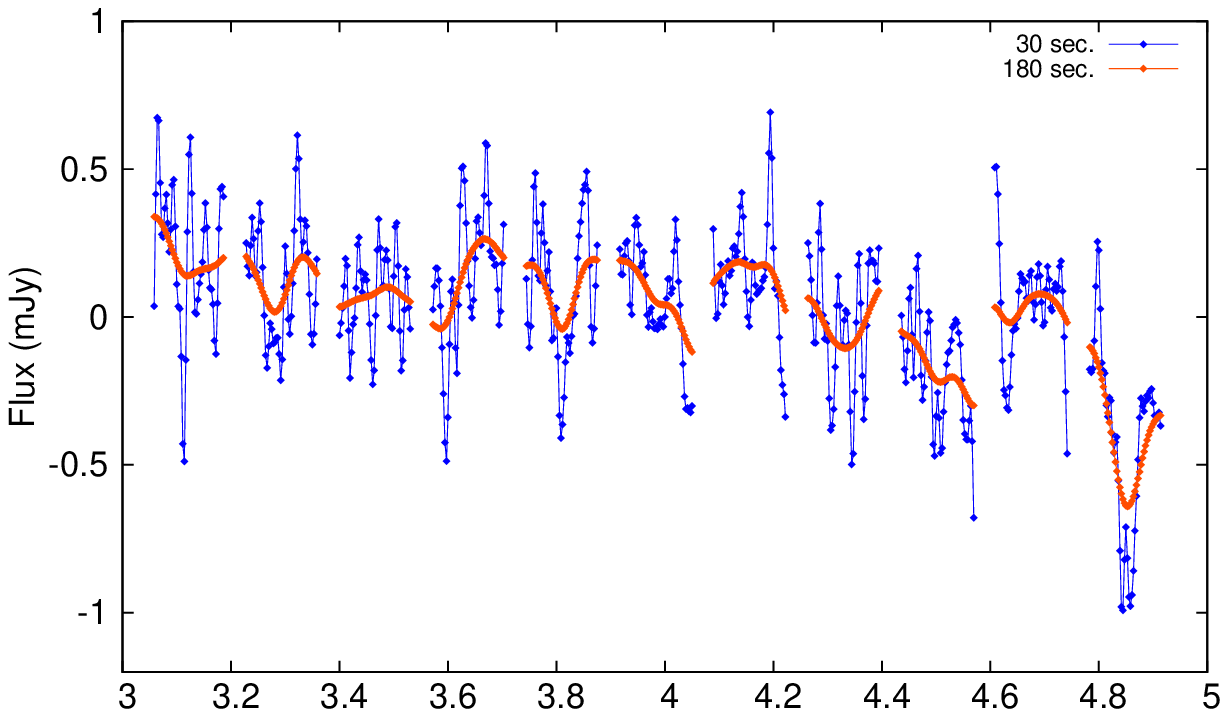} \vspace*{-1.5cm} \caption{ The light curves of
the total intensity (Stokes I) (left) and circular polarization
(Stokes V) (right) radio emission, detected at 4.9 GHz from
2MASS0746+20. The data, taken with the VLA in January 2007, is
smoothed over 10s (top panels) and  30 s \& 180 s (bottom
panels) time intervals.  The gaps in the data in the bottom panels of the
figure are due to phase calibration. In the Stokes V light curve, right circular
polarization is represented by positive values, while left circular
polarization is represented by negative values. A 100 $\%$ left
circularly polarized burst is evident in the last 10 minutes of the
observation. \label{Fig2}}
\end{figure*}

As a next step, we imaged the field around the dwarf separately in
three time intervals in both Stokes I and Stokes V, keeping a
close-by ($\sim$ 12 arcsec), bright background source for comparison.
Fig. 3 shows the resulting maps. The top left panel shows the Stokes
I map of the 3 to 4 hrs UT interval, the top right panel is the
Stokes I map during the 4 -- 4.75 hrs UT interval, while the bottom
left and right show Stokes I and Stokes V images of the last 10
minutes of the observation. The background source is the dominant
feature in the map during the first two intervals (as well as the
map of the entire 2 hour observation), while in the last one, it is
actually fainter than 2MASS 0746+20 in total intensity and is absent
from the map in circular polarization, where the ultracool dwarf is
clearly visible. Imaging the map in Stokes V during the first two
time intervals did not reveal a source at the position of the dwarf.
\begin{figure}[hbt!]
\vspace{9.0cm}
\includegraphics{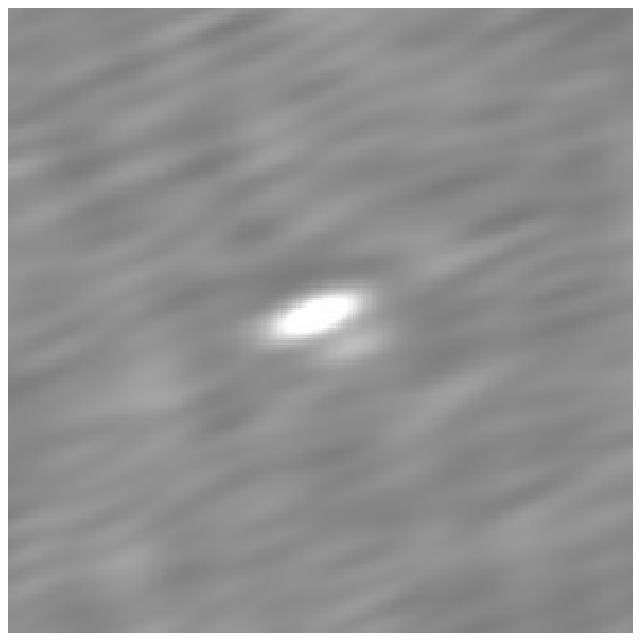}
\includegraphics{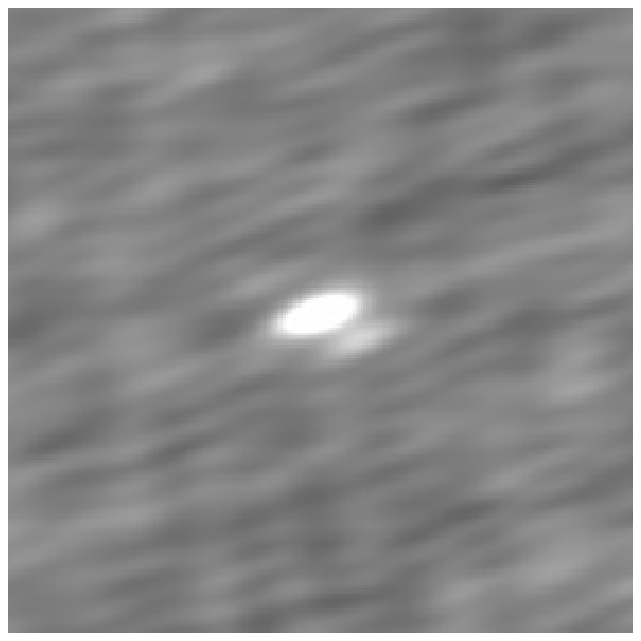}
\includegraphics{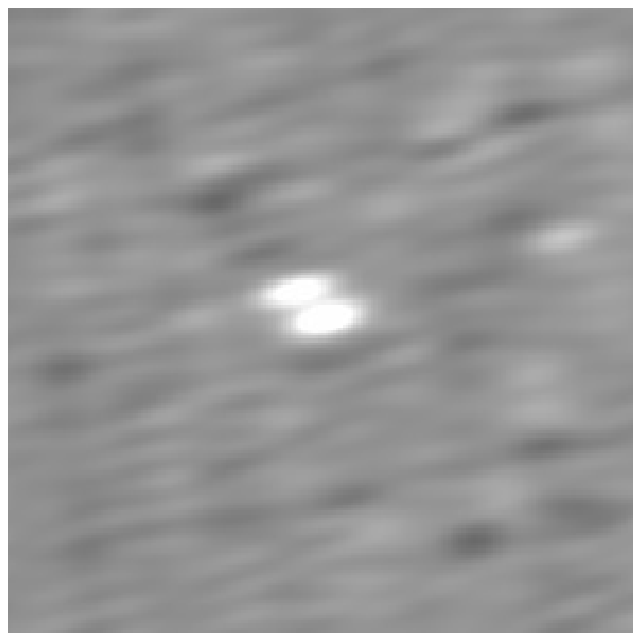}
\includegraphics{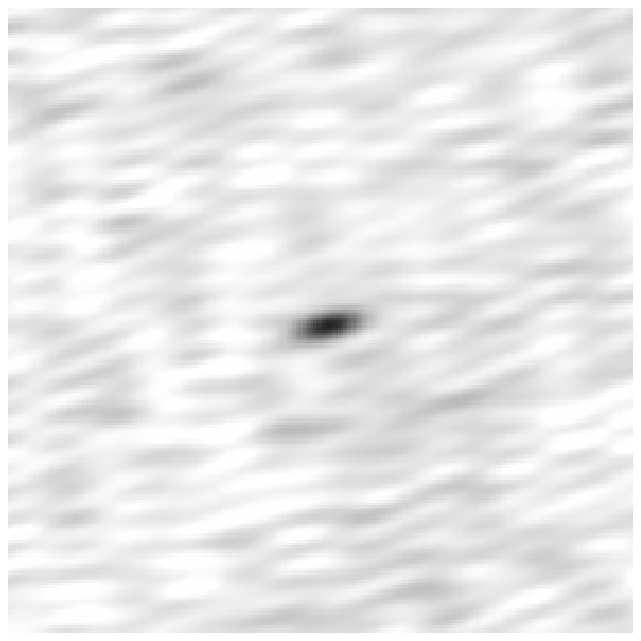}

\caption{Field maps of the binary dwarf 2MASS0746+20. Top left -
Total intensity (Stokes I) image of the first one hour of
observation (3 -- 4 hours UT). The dwarf's position is to the lower
right of the bright field source. Top right - Stokes I image of the
4 -- 4.75 hours UT interval, when the flux level of 2MASS0746+20 is
found to be rising. Bottom left - Stokes I image of the last 10
minutes of the observation. The dwarf is brighter than the
background source during this heightened period of emission. Bottom
right - Stokes V image of the last 10 minutes of the observation.
2MASS0746+20 is a source of highly circularly polarized emission,
while the background source is completely absent from the
map.\label{Fig3}}

\end{figure}

The behaviour of the 2MASS0746+20 light-curve, i.e. the low flux
levels at the start of the observation, followed by a brightening
and a strong, but short-lived burst, led us to an interesting
question. What if a source has pulsed emission, but its quiescent
emission is absent, or too low to be detected? In this case, since
the pulses span only a few minutes, the overall flux will be very
low and most likely at, or below the 3 $\sigma$ detection limit.
If $F_{\rm p}/F_{\rm q} > \sqrt(t_{\rm q}/t_{\rm p})$, where
$F_{\rm p}$ and $F_{\rm q}$ are the flux density during a pulse and
the average flux density of the whole observation, and $t_{\rm p}$
and $t_{\rm q}$ are the duration of the pulse and the duration of the
whole observation respectively, the SNR will be
higher for the pulse than for the entire observation. To
investigate such a possibility, we reexamined our X band data for
TVLM 513 (Hallinan \etal\ 2007), taken with the VLA in May 2006. It
spans over 10 hours, covers 5 periods of rotation of the dwarf and
clearly shows bursts of 100$\%$ left and right circularly polarized
flux. The SNR of a single 8 min burst exceeded the SNR of the entire
observation by factor of 1.5 (for 2MASS0746+20 the SNR during the
pulse exceeds the SNR of the whole observation by a factor of 2).
Therefore to rigorously determine whether an ultracool dwarf is detected
or not, it is necessary to image subsets of the data for the possible
presence of pulses.

To determine which subsets to image, we produced three sets of ten
second resolution time series of the May 2006 TVLM 513 data:
with the source at the phase center and the field sources removed,
with the source at the phase center and the field sources still present
and a set of time series with the source not at the phase center but with
the field sources present.
Comparing those three sets of time series, showed that even though
the presence of the background sources affects the shape of the
target's light-curve, the pulses are still clearly visible when the
target is at the phase centre of the map. If the target is shifted
from the phase centre, the shape of the light-curve changes
drastically and the pulses are no longer visible. We repeated the
same experiment on our 2MASS0746+20 data. The result was the same -
even with the background sources present, when the dwarf is at the
phase centre, i.e. the burst in the last 10 min of the observation is
still clearly present in the light curve. Shifting the position of
the target off, on the other hand, makes the picture less clear and
the presence of the burst is not obvious.

Encouraged by these results, we repeated the experiment for the rest
of the targets in our sample. We first acquired from the literature
accurate proper motion measurements for each dwarf, then calculated
the dwarfs' coordinates at the epoch of our observation. Using the
AIPS task UVFIX, we shifted the phase centre of every map to
coincide with the position of the respective dwarf and then plotted
the visibilities in both total intensity and circular polarization,
using the task DFTPL. Inspecting the light curves did not reveal
any additional detections.

\vspace*{-3mm}
\section{Discussion}
Until recently, radio emission from UCDs has been attributed to
incoherent gyrosynchrotron emission, similar to that detected from
early to mid-type M dwarfs (Berger \etal\ 2005, 2006, Burgasser \&
Putman 2005, Osten \etal\ 2006). However, Hallinan \etal\ (2006)
argued that the more likely scenario is a coherent, electron
cyclotron maser emission from a low-density region above the
magnetic poles of the dwarf. This model would require the presence
of a stable, large-scale magnetic field, with field strengths
$\approx$ 3 kG. The presence of fields with such strength on low
mass stars and UCDs has been suggested by direct observations
(Donati \etal\ 2007 and Reiners \& Basri 2007). Subsequent
observations of TVLM 513 (Hallinan \etal\ 2007) have revealed
extremely bright, periodic, 100$\%$ circularly polarized bursts,
produced by the strong beaming of the radio emission and the rapid
rotation of the dwarf. These properties, especially the high levels
of polarization and the inherent directivity of the emission,
together with the resulting high brightness temperature, provide
confirmation of the coherent nature of the emission.

For the electron cyclotron maser operation, a population inversion
in the electron distribution is needed, as well as a relatively
strong magnetic field and low-density plasma, so that the electron
cyclotron frequency $\nu{_c}$ is greater than the plasma frequency
$\nu{_p}$, where $\nu{_c} \approx 2.8 \times 10^6 {\rm B [Hz]}$
and $\nu{_p} \approx 9000 n_{e}^{1/2}$ Hz. An efficient
mechanism for reaching the necessary anisotropy in the electron
distribution, is the shell instability, proposed as a source of
Earth's auroral kilometric radiation, when it became clear that the
loss-cone distribution was a poor fit to the observations (see
Treumann 2006 for review). It is suggested that the main source for
any strong electron-cyclotron maser is found in the presence of a
magnetic-field-aligned electric potential drop which has several
effects. For example, it can dilute the local plasma to such an
extent that the plasma enters the regime in which the
electron-cyclotron maser becomes effective and favours emission in a
direction roughly perpendicular to the ambient magnetic field. This
emission is the most intense, since it implies the coherent resonant
contribution of a maximum number of electrons in the distribution
function. What is more, such an instability can be sustained over a
range of heights above the stellar surface, thus producing pseudo
broadband, coherent radio emission, which would explain the
simultaneous detection of both 8.4 and 4.9 GHz emission from TVLM
513 (Hallinan \etal\ 2006, 2007) and 2MASS0036+18 (Berger \etal\
2002, 2005). These conditions are fulfilled in the low density, high
magnetic field strength regions over the poles of a large scale
magnetic field.

The striking similarities in the emission of TVLM 513 and now
2MASS0746+20, in particular the highly polarized pulse, strongly
suggest that 2MASS0746+20 is an analog of TVLM 513, i.e. another
UCD producing electron cyclotron emission. Since we do not resolve
the system, we cansnot say which component is responsible
for the observed emission. Yet, as in the cases of TVLM 513, the
emission during the burst most likely comes from an active region
with a size much smaller than the radius of the dwarf.

ECM emission is emitted at the electron cyclotron frequency
$\nu_{c}$ and its harmonics, i.e. $\nu_{c} ~\approx$ 2.8 $\times
10^{6}$~B Hz. This gives maximum magnetic field strength
 B$_{{\rm max}} \approx$ 1.75 kG for 2MASS 0746+20 as compared to
 B$_{{\rm max}} \approx$ 3 kG derived for TVLM 513. Reiners \& Basri
(2007) recently suggested the presence of magnetic fields of such
strength on late M dwarfs via direct measurements, although these were
filling factor dependent. Of the fourteen dwarfs of spectral type M5.0 or
later, all but four show the presence of magnetic fields stronger
than 1 kG.

Further high sensitivity radio observations of 2MASS0746+20 are needed,
in particular a longer time series. Resolving the system to its
components will identify the component responsible for the radio
emission. It is also possible that both components are pulsing in
which case, using Fourier analysis techniques, one may be able to
extract the period of rotation of both dwarfs and determine the
inclination angle of the equatorial plane to the orbital plane,
using previously determined $v \sin i$.

\vspace*{-3mm}
\section{Conclusions}
The fraction of UCDs with detected radio emission so far is $\approx$
10\% (Berger 2006, Phan-Bao \etal\ 2006). Yet, recent observations
give us reason to think that this value may be underestimated. The
above figure is based on VLA surveys at 8.5 GHz and with the
assumption that the emission from these objects is stable over long
periods of time. However, in the context of the electron cyclotron
maser, emission at 8.5 GHz would require the presence of magnetic
fields strengths up to $\approx$ 3 kG. Thus any objects with weaker
fields would be undetected at that frequency, as is the case of
2MASS0746+20. On the other hand, recent VLA observations of the
L2.5 brown dwarf 2MASS J05233822-1403022 reveal the possibility that
levels of ECM emission generated in UCDs can vary significantly over
months (Antonova \etal\ 2007).

Taking into account the above considerations, it is clear that
despite the progress made in understanding the production mechanism
of radio emission from ultracool dwarfs, the actual fraction of
dwarfs, capable of producing radio emission, as well as the fraction
of those, that show periodic pulsations is still unclear. The
detection of emission at 4.9 GHz from 2MASS0746+20 points strongly
towards the electron cyclotron maser instability as a more likely
emission mechanism, at least for the short duration pulse. However,
further followup observations are required. \\

{\bf Acknowledgements}
Armagh Observatory is grant-aided by the N. Ireland Dept. of
Culture, Arts \& Leisure. We gratefully acknowledge the support
of Science Foundation Ireland (Grant Number 07/RFP/PHYF553). This
research has made use of the Simbad database, operated at CDS,
Strasbourg, France.

\end{document}